\documentclass[twocolumn,aps,showpacs,preprintnumbers,nofootinbib,
superscriptaddress]{revtex4}
\usepackage{graphicx}
\usepackage{dcolumn}
\usepackage{bm}
\begin{document}


\title{Enhanced radiative ion cooling}

\author{E.G.Bessonov, Lebedev Physical Institute RAS, Moscow, Russia}

\date{\today}

                       \begin{abstract}
Enhanced laser cooling of ion beams and Robinson's damping criterion
are discussed.
                       \end{abstract}


\maketitle
           \section{Introduction}

In the method of ordinary three-dimensional radiative cooling of ion
beams a laser beam overlaps an ion beam, its transverse position is
motionless, all ions interact with the laser beam independent of their
energy and amplitude of betatron oscillations. The difference in rates
of momentum loss of ions having maximum and minimum energies in the
beam is small. That is why the cooling time of the ion beam is high
\cite {idea} - \cite {pac}.

When a cooling is produced in a dispersion-free straight section and
the laser beam intensity is constant inside the area of the laser beam
occupied by a being cooled ion beam, then the damping times of the
horizontal vertical and phase oscillations are:

        \begin{equation}   
        \tau _x = \tau _y = {\tau _{\epsilon} \over 1 + D} =
        {2\varepsilon \over \overline P},
        \end{equation}
where $\overline P$ is the average power of scattered radiation;
$\varepsilon$, ion energy; $D$, saturation parameter.

The physics of radiative ion cooling is similar to synchrotron
radiation damping. Friction causes the appearance of a reaction force
on the emitting particle. Liouville's theorem does not valid for such
non-conservative system. At the same time in the case of non-selective
interaction the Robinson's damping criterion is valid:  the sum of
damping decrements $\tau _{x,y,s} ^{-1}$ is a constant. In a particular
choice of lattice and laser or material target, damping rates can be
shifted between different degrees of freedom. However the decrements
are limited by the criterion by the value $\tau _{x,y,s} \geq 2(1 + D)
\varepsilon /(3 + D) \overline P$ when all decrements are greater then
or equal to zero \cite {wiedemann, ICFA01}.

                \section {Enhanced laser cooling}

The method of enhanced resonance laser cooling of ion
beams in the longitudinal plane is based on the Rayleigh scattering of
"monochromatic" laser photons by not fully stripped ion beams or by
complicated nuclei when the radio-frequency (RF) system of the storage
ring is switched off \cite {channel} - \cite {hangst1}.  In this method
the laser beam overlaps the ion beam and has a chirp of frequency. Ions
interact with the laser beam at resonance energy, decrease their energy
in the process of the laser frequency scanning until all of them reach
the minimum energy of ions in the beam. At this frequency the laser
beam is switched off. The higher energy of ions the earlier they begin
interaction with the laser beam, the longer the time of interaction.
Ions of minimum energy do not interact with the laser beam at all. In
such a way, in this method, the selective interaction is realized. The
damping time of the ion beam in the longitudinal plane is determined by
the dispersion $\sigma _{\varepsilon}$ of its energy spread

        \begin{equation}   
        \tau _{\epsilon} = {2 \sigma _{\varepsilon} \over
        \overline P}. \end{equation}

The damping time (2) is $\varepsilon / \sigma _{\varepsilon} \sim 10
^3$ times less than (1). At the same time the transverse decrements are
zero. From this it follows that the selectivity of interaction leads to
the violation of the Robinson's damping criterion and open the
possibility of the enhanced cooling.

Below we will consider the non-resonance methods of enhanced cooling of
ion beams in the longitudinal and transverse planes. These methods can
be used for cooling of another particles as well. A universal kind of
selective interaction of particles with moving in the radial direction
laser or media targets will be used. By analogy with the resonance
laser cooling the interaction region in this case change its radial
position in the being cooled beam for the cooling process when the
dispersion function of the storage ring is not equal to zero at the
interaction region. Robinson's damping criterion does not work in this
case as well (see Appendix).

          \subsection {New enhanced cooling methods}

For the sake of simplicity we will neglect the emission of the
synchrotron radiation by ions in the bending magnets of a storage ring,
supposing that the RF system of the ring is switched off and the
broadband laser beams (or material targets) are homogeneous and have
sharp edges in the radial directions. We suppose that the jump of
instantaneous orbits of ions caused by the energy loss is less than the
amplitude of their betatron oscillations.

In a smooth approximation, the motion of an ion relative to its
instantaneous orbit is described by the equation $x _{\beta} = $
$A\cos(\Omega t+\varphi)$, where $x _{\beta} = x - x_{\eta}$ is the ion
deviation from the instantaneous orbit $x_{\eta}$; $x$, its radial
coordinate; $A$ and $\Omega$, the amplitude and the frequency of
betatron oscillations. If the coordinate $x _{\beta\,0}$ and transverse
radial velocity of the ion $\dot x _{\beta\,0} = - A \Omega \sin
(\Omega t _0$ $ + \varphi )$ correspond to the moment $t _0$ of change
of the ion energy in a laser beam then the amplitude of betatron
oscillations of the ion before an interaction is $A _0$ $ =$ $ \sqrt {x
_{\beta\,0} ^2 + \dot x _{\beta \,0} ^2 /\Omega ^2}$. After the
interaction, the position of the ion instantaneous orbit will be
changed by a value $\delta x_{\eta}$, the deviation of the ion relative
to the new orbit will be $x _{\beta \,0} - \delta x _{\eta}$, and the
direction of the electron velocity will not be changed. The new
amplitude will be $A_1$ $ =\sqrt {(x _{\beta \,0} - \delta x _{\eta})^2
+ \dot x ^2 _{\beta \,0} /\Omega ^2}$ and the change of the square of
the amplitude

       \begin{equation}
       \delta (A)^2 = A_1^2 - A_0^2 = - 2x _{\beta \,0}\delta x _{\eta}
       + (\delta x _{\eta}) ^2.
       \end{equation}          

When $|\delta x _{\eta}| \ll |x _{\beta\,0}| < A _0$ then in the first
approximation the value $\delta A = - (x _{\beta \,0} /A) \delta x
_{\eta}$. From this it follows that to produce the enhanced cooling of
an ion beam in the transverse plane we must create such conditions when
ions interact with a laser beam under deviations from the instantaneous
orbit $x _{\beta 0}$ of one sign. In this case the value $\delta A $
has one sign and the rate of change of amplitudes of betatron
oscillations of ions is maximum.  A selective interaction of ions with
the laser beam is necessary to realize this case.

\vskip 0mm
\begin{figure}[hbt]
\includegraphics{
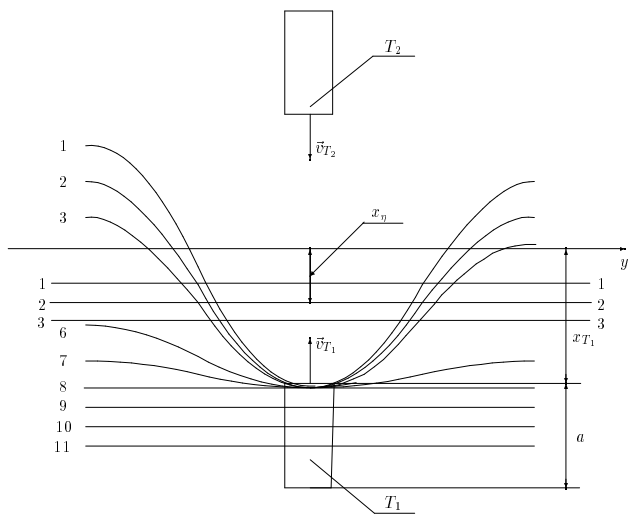}
\caption{\label{fig:epsart} The scheme of the enhanced electron
cooling. \\The axis $"y"$ is the equilibrium orbit of the storage ring;
$T_1$ and $T_2$ the laser beams. The transverse positions of laser beams
are displaced with the velocity $\vec v _{T _{1,2}}$ relative to the
equilibrium orbit, 1-1, 2-2, ...  the location of the instantaneous
electron orbit, and 1,2,3, ... the electron trajectories after 1,2,3,
...  events of the energy loss.}
\end{figure}
\vskip 2mm

Two schemes of a selective interaction of ion and laser beams for
cooling of ion beams in the transverse and longitudinal planes can be
suggested (see Fig.1 ). For the transverse cooling, the laser beam $T
_1$ is used. At the initial moment it overlaps a small external part
of the ion beam in the radial direction in the straight section of the
storage ring with non zero dispersion function. First ions with largest
initial amplitudes of betatron oscillations interact with the laser
beam. Immediately after the interaction and loss of the energy the
position and direction of momentum of an ion remain the same, but the
instantaneous orbit is displaced inward in the direction of the laser
beam. The radial coordinate of the instantaneous orbit and the
amplitude of betatron oscillations are decreased to the same value
owing to the dispersion coupling. After every interaction the position
of the instantaneous orbit approaches the laser beam more and more, and
the amplitude of betatron oscillations is coming smaller. It will reach
some small value when the instantaneous orbit reaches the edge of the
laser beam.  When the depth of dipping of the instantaneous orbit of
the ion in the laser beam becomes greater than the amplitude of its
betatron oscillations, the orbit will continue its motion in the
laser beam with constant velocity. The amplitude of betatron
oscillations will not be changed.

The degree of overlapping is changed by moving uniformly the laser beam
position from inside in the direction of the being cooled ion beam
with some velocity $v _{T _1}$\footnote {Instantaneous orbits can be
moved in the direction of the laser beam, instead of moving of a laser
beam.  A kick, decreasing of the value of the magnetic field in bending
magnets of the storage ring, a phase displacement or eddy electric
fields can be used for this purpose.}.  When the laser beam reaches the
instantaneous orbit corresponding to ions of maximum energies then
the laser beam must be switched off and returned to a previous
position. All ions of the beam will have small amplitudes of
betatron oscillations and increased energy spread. Ions with high
amplitudes of betatron oscillations will start to interact with a laser
beam first, their duration of interaction and absolute decrease of
amplitudes of betatron oscillations will be higher.

To realize the enhanced cooling of an ion beam in the longitudinal
plane we can use a broadband laser beam $T _2$ located in the region of
a storage ring with non zero dispersion function. The radial laser beam
position is moving uniformly from outside in the direction of the being
cooled ion beam with a velocity $v _{T _2}$ higher than maximum
velocity $\dot x_{\eta \, in}$ of the ion instantaneous orbit deepened
in the laser beam. At the initial moment, the laser beam overlaps only
a small part of the ion beam. The degree of overlapping is changed in
such a way that ions of maximum energy, first and then ions of lesser
energy, come into interaction. When the laser beam reaches the orbit of
ions of minimum energy then it must be switched off and returned to the
previous position. In this case, the rate of the energy loss of ions in
the beam will not be increased, but the difference in duration of
interaction and hence in the energy losses of ions having maximum and
minimum energies will be increased essentially.  As a result all ions
will be gathered at the minimum energy in a short time.

      \subsection {Interaction of ion beams with transversely
      moving laser beams}

In the methods of enhanced laser cooling of ion beams the internal
and external laser beam positions are displaced in the transverse
directions. Below the evolution of amplitudes of betatron oscillations
and positions of instantaneous orbits in the process of the energy loss
of ions in laser beams will be analyzed.

The velocity of an ion instantaneous orbit $\dot x_{\eta}$ depends
on the distance $x _{T _{1,2}} - x _{\eta}$ between the edge of the
laser beam and the instantaneous orbit, and on the amplitude of
betatron oscillations. When the orbit enters the laser beam at the
depth higher than the amplitude of betatron oscillations then ions
interact with the laser beam every turn and theirs velocity reaches the
maximum value $\dot x_{\eta \, in}$ which is given by the intensity and
the length of the interaction region of the ion and laser beams.
In the general case, the velocity $\dot x_{\eta}$ can be presented in
the form $\dot x_{\eta} =  W\cdot \dot x_{\eta \, in}$, where $W$ is
the probability of an ion crossing the laser beam. $W$ is the
ratio to a period of a part of the period of betatron oscillations of
the ion determined by the condition $| x _{T _{1 2}} - x_{\eta}| \leq
|x _{0}| \leq A$ when the deviation of the ion from the instantaneous
orbit is directed to the laser beam and is greater than the distance
between the orbit and the laser beam. Probability can be presented in
the form $W = \varphi _{1, 2}/\pi$, where $\varphi _{1} = \pi - \arccos
\xi _{1}$, $\varphi _{2} = \arccos \xi _{2}$, $\xi _{1, 2} = (x _{T_{1,
2}} - x _{\eta}) /A$, indices 1,2 correspond to laser beams.

The behavior of the amplitudes of betatron oscillations of ions,
according to (3), is determined by the equation $\partial A/\partial x
_{\eta} = -<x _{\beta\,0}>/A$, where $<x _{\beta\,0}>$ is the ion
deviation from the instantaneous orbit averaged through the range of
phases $2 \varphi _{1, 2}$ of betatron oscillations where ions
cross the laser beam. The value $<x _{\beta\,0}> = \pm A sinc \,
\varphi _{1 ,2}$, where $sinc \varphi _{1, 2} = sin \varphi _{1,
2}/\varphi _{1, 2}$, signs $+$ and $-$ are related to the first and
second laser beams. Thus the cooling processes are determined by the
system of equations

       \begin{equation}               
       {\partial A\over \partial x_{\eta}} = \pm sinc \varphi
       _{1,2}, \hskip 5mm {\partial x _{\eta}\over \partial t} = {\dot
       x_{\eta \, in} \over \pi}\varphi _{1,2}.
       \end{equation}

From equations (4) and the expression $\partial A/\partial x _{\eta} =
[\partial A/\partial t]/ [\partial x _{\eta}/\partial t]$ it follows:

       \begin{equation}                                
       {\partial A\over \partial t} = {\dot x _{\eta \,in}\over
       \pi}\sin\varphi _{1,2} = {\dot x _{\eta \,in}\over
       \pi}\sqrt {1 - \xi _{1,\,2}^2}.  \end{equation}

Let the initial instantaneous ion orbits be distributed in a
region $\pm \sigma _{x, \varepsilon, 0}$ relative to the location of
the middle instantaneous orbit $\overline {x _{\eta}}$, and the initial
amplitudes of ion radial betatron oscillations $A _0$ be
distributed in a region $\sigma _{x,b,0}$ relative to their
instantaneous orbits, where $\sigma _{x, \varepsilon, 0}$ and $\sigma
_{x,b,0}$ are dispersions. The dispersion $\sigma _{x, \varepsilon, 0}$
is determined by the initial energy spread $\sigma _{\varepsilon, 0}$.

Suppose that the initial spread of amplitudes of betatron oscillations
of ions $\sigma _{x,b,0}$ is identical for all instantaneous orbits of
the beam. The velocities of the instantaneous orbits in a laser beam
$\dot x _{\eta \, in} < 0$, the transverse velocities of the laser
beams $v _{T _{1}} > 0$, $v _{T _{2}} < 0$. Below we will use the
relative radial velocities of the laser beam displacement $k _{1,2} = v
_{T _{1,2}}/\dot x _{\eta \, in}$, where $v _{T _{1,2}} = dx _{T
_{1,2}}/dt$.  In our case $k _1 < 0$, $k _2 > 0$.

From the definition of $\xi _{1, 2}$ we have a relation $x _{\eta} = x
_{T _{1,2}} - \xi _{1, 2} A(\xi _{1, 2})$. The time derivative is
$\partial x _{\eta}/\partial t = v _{T _{1,2}} - [A + \xi _{1,2}
(\partial A/ \partial \xi _{1,2})] \partial \xi _{1,2} /\partial t$.
Equating this value to the second term in (4) we will receive the time
derivative

        \begin{equation}                               
        {\partial \xi _{1,2}\over \partial t} = {\dot x _{\eta \,in}
        \over \pi} {\pi k _{1,2} - \varphi _{1, 2} \over A(\xi
        _{1,2}) + \xi _{1,2} (\partial A / \partial \xi
        _{1,2})}.  \end{equation}

Using this equation we can transform the first value in (4) to the form
$\pm sinc \varphi _{1,2} (\xi _{1,2}) = ({\partial A/ \partial \xi _{1,
2}}) ({\partial \xi _{1,2}/ \partial t})/$ $ ({\partial x _{\eta}/
\partial t}) = $ $(\pi k _{1,2} - \varphi _{1,2}) {(\partial A
/\partial \xi _{1,2})}/ [A + \xi _{1,2}(\partial A/$ $\partial \xi
_{1,2})]\cdot $ $\varphi _{1,2}$ which can be transformed to $\partial
\ln A / \partial \xi _{1,2} = \pm \sin \varphi _{1,2} / \pi k _{1,2} -
(\varphi _{1,2} \pm \xi _{1,2} \sin \varphi _{1,2}).$ The solution of
this equation is

       \begin{equation}                         
       A = A _0 \exp \int _{\xi _{1,2,0}} ^   {\xi _{1,2}}
       {\pm \sin \varphi _{1,2} d\xi _{1,2}
       \over \pi k _{1,2} - (\varphi _{1,2}\pm \xi _{1,2}
       \sin \varphi _{1,2})},
       \end{equation}
where the index $0$ correspond to the initial time. Substituting the
values $A$ and $\partial A/ \partial \xi _{1,2}$ determined by (7) in
(6) we find the relation between time of observation and parameter
$\xi _{1,2}$

       \begin{equation}                                
       t - t _0 = {\pi A _0 \over |\dot x _{\eta \,in}|} \psi
       (k _{1,2}, \xi _{1,2}),   \end{equation}
where $\psi (k _{1,2}, \xi _{1,2}) = - \int ^{\xi _{1,2 }}
_{\xi _{1,2,  0}} A(\xi _{1,2 })/ \{A_0 [ \pi k _{1, 2} - $
$ (\varphi _{1,2} \pm \xi _{1,2 }\sin \Delta \varphi
_{1,2})]\}d\xi _{1,2 }. $

The equations (8) determine the time dependence of the functions
$\xi _{1,2}(t  - t _0)$. The dependence of the amplitudes
$A[\xi _{1,2} (t - t _0)]$ is determined by the equation (7)
through the functions $\xi _{1,2}(t - t_0)$ in a parametric form. The
dependence of the position of the instantaneous orbit follows from
the definition of $\xi _{1,2}$

       $$x_{\eta}(t - t _0) = x_{T _{1, 2}0} + v _{T_{1,2}}(t  -
       t_0) - $$
\vskip -7mm
       \begin{equation}                            
       A[(\xi _{1, 2} (t  - t_0)] \cdot \xi (t  - t _0).
       \end{equation}

The function $\psi (k _2,\xi _{2})$ for the case $k _2 > 0$ according
to (8) can be presented in the form

       $$\psi (k _2, \xi _{2}) = \int _{\xi _{2}} ^{1}{dx\exp
       \int ^{1} _{x}} $$

       \begin{equation}             
       {{\sqrt{1 - t^2} / (\pi k _{2} - \arccos t +
       t\sqrt {1 - t ^2})} \over \pi k _{2} - \arccos x + x \sqrt
       {1 - x ^2}}d\,t.
       \end{equation}

The instantaneous orbits of ions having initial amplitudes of
betatron oscillations $A _0$ will be deepened into the laser beam to
the depth greater than their final amplitudes of betatron oscillations
$A _f$ at a moment $t _f$.  According to (8), $t _f = t _0 + \pi A _0
\psi (k _2, \xi _{2,f})/|x _{\eta \, in}|$, where $\xi _{2,f} = \xi (t
_f) = 1$. During the interval $t _f - t _0$ the laser beam $T _2$ will
pass a way $l _f = |v _{T _2}|(t _f - t _0) = \pi k_2 \psi (k _2, \xi
_{2,f}) A _0$.  The dependence $\psi (k _2, \xi _{2, f})$ determined by
(10) is presented in Table 1.

\vskip 2mm
\begin{figure}[hbt]
Table 1
\hskip 20 mm \vskip 2mm
\begin{tabular}{|l|l|l|l|l|l|l|l|l|l|l|l|l|l|}
\hline
$k _2$ & 1.0 & 1.02 & 1.03 & 1.05 & 1.1&1.2 &1.3&1.4&1.5&1.7&2.0 \\
\hline
$\psi $
&$\infty$&13.8&9.90&6.52&3.71&2.10&1.51&1.18&.98& .735&.538\\
\hline
\end{tabular}
\end{figure}

\vskip 2mm

Numerical calculations of the dependence $\psi (k _2, \xi _{2})$ on
$\xi _{2}$ for the cases $k _2 = 1.0$, $k _2 = 1.1$ and $k _2 = 1.5$
are presented in Tables 2, 3, and 4, respectively. It can
be presented in the next approximate form

       \begin{equation}         
       \psi (k _2, \xi _{2}) \simeq C _3(k _2)\psi ({1 - \xi _{2}
       \over k _2 + \xi _{2}}),
       \end{equation}
where $C _3(k _2) \simeq 0.492 - 0.680 (k _2 -1) + 0.484(k _2 - 1)^2 +
...$, $\psi [(1 - \xi _{2}) / (k _2 + \xi _{2})]|_{k _2 = 1}
\simeq (1 - \xi _{2}) / (1 + \xi _{2})$.

\vskip4mm
\begin{figure}[hbt]
Table 2 \hskip 20 mm ($k_2 = 1.0$)
\vskip 2mm
\begin{tabular}{|l|l|l|l|l|l|l|l|l|l|l|}
\hline
$\xi _{2}$ & 1.0 & 0.5 & 0.2& 0&-0.2&-0.5&-0.8&-0.9& -1.0 \\
\hline
$\psi $ & $0$ &.182&.341&.492&.716& 1.393&4.388 &10.187&-$ \infty $\\
\hline \end{tabular}

\vskip 10mm
Table 3 \hskip 20 mm ($k_2 = 1.1$)
\vskip 2mm
\begin{tabular}{|l|l|l|l|l|l|l|l|l|l|l|}
\hline
$\xi _{2}$ & 1.0 & 0.5 & 0.2&0 &-0.2&-0.5& -0.8 & -0.9& -1.0  \\
\hline
$\psi $&$0$&.163&.300&.423&.595&1.033& 2.076&2.759
& 3.710 \\
\hline \end{tabular}

\vskip 10mm
Table 4  \hskip 20 mm ($k_2 = 1.5$)
\vskip 2mm
\begin{tabular}{|l|l|l|l|l|l|l|l|l|l|l|l|l|l|}
\hline
$\xi _{2}$ & 1.0 & 0.5 & 0.2&0 & -0.2& -0.4&-0.6&-0.8&-1.0  \\
\hline
$\psi $&$0$&0.116&0.202&0.273&0.359& 0.466 &0.602 &0.772 &0.980 \\
\hline
\end{tabular}
\end{figure}
\vskip 10mm

\subsection {The enhanced transverse laser cooling}

In the method of the enhanced transverse laser cooling of ion
beams a laser beam $T_1$ is located in the region ($x _{T _1}$, $x_{T
_1} - a$), where $a$ is the laser beam width. The degree of the
transverse cooling of the ion beam is determined by (7). The final
amplitude in this case can be presented in the form

       \begin{equation}           
       A _f = A _0 exp {\int _{\xi _{1,0}} ^{\xi _{1,f}}
       { \sqrt{1 - \xi _1^2} d\xi _1 \over \pi k _{1} - \pi +
       \arccos \xi _1 - \xi _1 \sqrt {1 - \xi _1 ^2}}}.
       \end{equation}

The numerical calculations of the dependence of the ratio $A _f /A _0$
on the relative radial velocity $k _1$ of the laser beam displacement
are presented in Fig.2 and Table 5 for the case $\xi _{1,0} = - 1$,
$\xi _{1,f} = 1$. This dependence can be presented by the approximate
expression

       \begin{equation}          
       A _{f} \simeq A _0 \sqrt{|k_1|\over |k _1| +1}.
        \end{equation}

\begin{figure}[hbt]
\includegraphics{
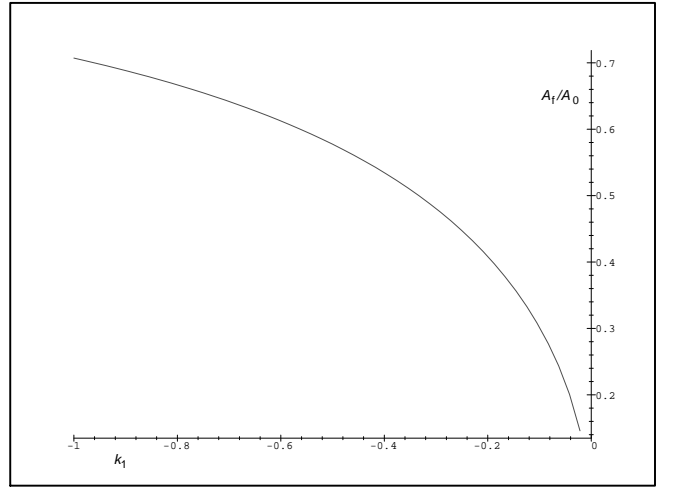}
\caption{\small \it The dependence of the ratio $A _{f}/A_0$ on $k _1$.}
\end{figure}
\vskip 4mm

\begin{figure}[hbt]
Table 5
\vskip 3mm
\begin{tabular}{|l|l|l|l|l|l|l|}
\hline
$|k_1|$ &0&0.2&0.4&0.6&0.8&1.0\\
\hline
$A_{f}/A_0$  &0&0.408&0.535&0.612&0.667&0.707\\
\hline
\end{tabular}
\end{figure}
\vskip 3mm

The time of the laser beam cooling and the final total radial dimension
of the beam are equal to

         \begin{equation}                
         \tau _{x,1} \simeq {\sigma _{x, 0} \over v _{T _1}},
         \hskip 5mm \sigma _{x,f}|_{|k _1| \ll 1} \simeq {\sigma _{x,0}
         \over |k _1|} + \sigma _{x, \varepsilon, 0},
         \end{equation}
where $\sigma _{x,0} = \sigma _{x,b,0} + \sigma _{x, \varepsilon,0}$ is
the total initial radial dimension of the ion beam. For the time
$\tau _{x,1}$ the instantaneous orbits of ions of a beam having
minimum energy and maximum amplitudes of betatron oscillations at $|k
_1| \ll 1$ pass the distance $\sim |\dot x _{\eta\, in}| \tau _{x,1}$.

According to (13) and (14) the enhanced transverse laser cooling can
lead to an appreciable degree of cooling of ion beams in the transverse
plane and a much greater degree of heating in the longitudinal one.

\subsection {The enhanced longitudinal laser cooling}

In the method of the enhanced longitudinal laser cooling of ion
beams a laser beam $T _2$ is located in the region ($x _{T _2}$, $x_{T
_2} + a$). Its radial position is displaced uniformly
with the velocity $v _{T _2} < 0$, $|v _{T _2}| > |\dot x _{\eta \, in}|$
from outside of the working region of the storage ring in the direction
of a being cooled ion beam. The instantaneous orbits of ions
will go in the same direction with a velocity $|\dot x _{\eta}| \leq
|\dot x _{\eta \, in}|$ beginning from the moment of their first
interaction with the laser beam. When the laser beam reaches the
instantaneous orbit of ions having minimum initial energies it
must be removed to the initial position.

The law of change of the amplitudes of ion betatron oscillations
is determined by (7), which can be presented in the form

       \begin{equation}                                 
       A  = A _0 exp {\int _{\xi _{2, 0}} ^{\xi _{2}}{- \sqrt{1
       - \xi _2^2} d\xi _2 \over \pi k _{2} - \arccos \xi _2+ \xi
       _2\sqrt {1 - \xi _2^2}}}.  \end{equation}

\vskip 30mm
\begin{figure}[hbt]
\includegraphics{
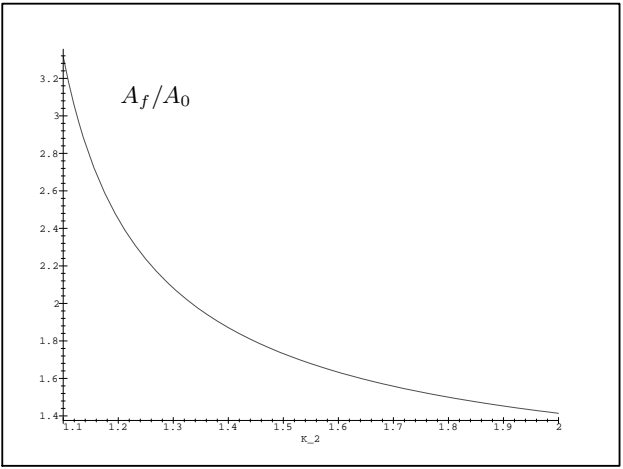}
\caption{\small \it The dependence of the ratio $A _{f}/A_0$ on $k
_2$.}
\vskip -60mm
\hskip -40mm
{\small $A _{f}/A_0$}
\end{figure}
\vskip 45mm

\vskip 2mm
\begin{figure}[hbt]
\vskip 6mm
Table 6
\vskip 3mm
\begin{tabular}{|l|l|l|l|l|l|l|}
\hline
$k_2$ & 1.0001 & 1.001& 1.01 & 1.1& 1.5 & 2.0 \\
\hline
$A _{f}/A _0$ &$100.005$&31.64&10.04 &3.32& 1.73&1.414 \\
\hline
\end{tabular}
\end{figure}

The dependence of the ratio of a final amplitude of ion betatron
oscillations $A _f = A(\xi _2 = 1)$ to the initial one on the relative
velocity $k _2$ of the second laser beam is presented in Fig.3 and
Table 6. This ratio can be presented by the next approximate
expression

       \begin{equation}                            
       A _{f} \simeq  A _0\sqrt {k_2\over k_2 - 1}.
        \end{equation}

The evolution of instantaneous orbits of ions interacting with the
laser beam depends on the initial amplitudes of betatron oscillations
of these ions. First of all the laser beam $T _2$ interacts with
ions having the largest initial amplitudes of betatron
oscillations $A _0 = \sigma _{x,b,0}$ and the highest energies. The
instantaneous orbit of these ions, according to (7) - (9), is
changed by the law $ x _{\eta _1} = x _{T _2, 0} + v _{T _2}(t - t _{0
_1}) - \xi _{2} \sigma _{x,b}(\xi _{2})$ up to the time $t = t _f$,
where $t _{0 _1}$ is the initial time of interaction of ions with
the laser beam. At the same time instantaneous orbits $x _{\eta _2}$ of
ions having the same maximum energy but zero amplitudes of
betatron oscillations are at rest up to the moment $t _{0 _2} = t _{0
_1} + \sigma _{x,b,0}/|v _{T_2}|$. The orbit $x _{\eta _1}$ is
displaced relative to the orbit $x _{\eta _2}$ by the distance $\Delta
x _{\eta _{1-2}} = (x _{\eta _1} - x _{\eta _2})$. At the moment $t _{0
_2}$, when $x _{\eta _2} = x _{T _2}$, this distance reaches the minimum

       \begin{equation}
       \Delta x _{\eta _{1-2} m}(t _{0 _2}) =
       - \xi _2(t _{0 _2})\cdot \sigma _{x,b}(t _{0 _2}) < 0,
       \end{equation} 
where the parameter $\xi _2(t _{0 _2})$, according to (8) and the
condition $|v _{T _2}|(t _{0 _2} - t _{0 _1}) = \sigma _{x,b,0}$, will
be determined by the equation $\psi [k _2, \xi _{2}(t _{0 _2})] = 1/\pi
k_2$. The value $\psi [k _2, \xi _{2}(t _{0 _2})]| _{k_2 \simeq 1}
\simeq 1/\pi$, $\xi _{2} (k _2, t _{0 _2})|_{k _2 \simeq 1}$ $ \simeq
0.22$ (see Tables 2-4), $\sigma _{x,b}(t _{0 _2}) = 1.26 \sigma
_{x,b,0}$ and the distance $|\Delta x _{\eta _{1-2}}(t _{0 _2})| \simeq
0.28 \sigma _{x,b,0}$. This distance is decreased with increasing $k
_2$.

The instantaneous orbit of particles $x _{\eta _2}$ inside the interval
$t _{0 _2} < t \leq t _f$ is changed by the law $x _{\eta _2} = x _{T
_2, 0}$ $ - \sigma _{x,b,0} + \dot x _{\eta,\,in}(t - t _{0 _1} +
\sigma _{x,b,0}/v _{T_2})$ and the distance

           $$\Delta x _{\eta _{1- 2}} = {(k _2 -1)\over k _2}
          [\sigma _{x,b,0} + v _{T _2}(t - t _{0 _1})] -
          \xi _{2} \sigma _{x, b}(\xi _{2}) = $$
          \begin{equation}    
           - [{k _2 - 1\over k _2} ({l _{T _2}\over \sigma _{x,b,0}}
           - 1) + \xi_{2} D _{2})]\sigma _{x,b,0},
       \end{equation}
where $D _{2} = D _2(k _2, \xi _{2}) =  \sigma _{x,b}/\sigma _{x,b,0}
= A / A _0$ , $l _{T _2} = x _{T _2} - x _{T _2, 0} = \pi k_2 \psi (k
_2, \xi _{2}) \sigma _{x,b,0} \leq l _f$ is the displacement of the
laser beam. The typical dependence $D _{2}$ defined by (15) is
presented in Fig.4.

\vskip 25mm

\begin{figure}[hbt]
\includegraphics{
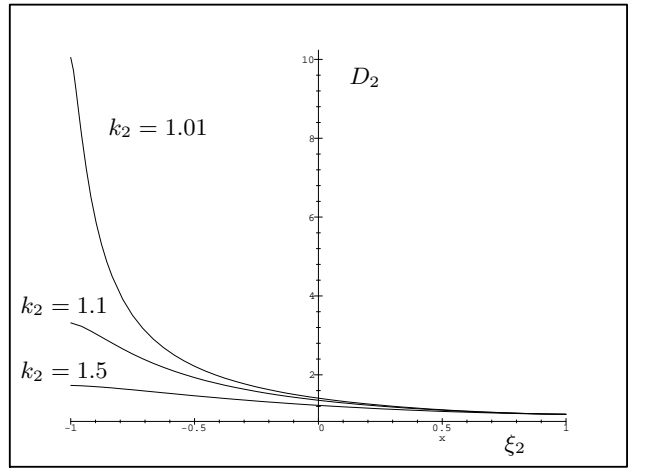}
\caption{\small \it The dependence on $\xi _{2}$ of the ratio of a
current amplitude of betatron oscillations to an initial amplitude $ D
_{2} = A /A _0 = \sigma _{x,b}/ \sigma _{x,b,0}$.}

\vskip -70mm
\hskip 10mm
{\small $D _{2}$}

\vskip 3mm
\hskip -45mm
{\small $k _2 = 1.01$}

\vskip 20mm
\hskip -70mm
{\small $k _2 = 1.1$}

\vskip 5mm
\hskip -70mm
{\small $k _2 = 1.5$}

\vskip 6mm
\hskip 50mm
{\small $\xi _{2}$}

\end{figure}
\vskip 25mm

When $t > t _f$ then the value $\xi _{2} = \xi _{2,f} = -1$, $l _{T _2}
= l _f$, $D _{2} = \sqrt {k _2/(k _2 - 1)}$ and (18) have the maximum

       $$\Delta x _{\eta _{1-2}}| _{t > t _f}  =
       [{k _2 - 1\over k _2} + \sqrt {k _2\over k _2 -1} -$$
       \begin{equation}  
       \pi (k _2 -1) \psi (k _2, \xi _{2,f})
       ]\sigma _{x,b,0}.
       \end{equation}

The instantaneous orbit $ x _{\eta _{2}}$ will be at a distance $x
_{\eta _{2-3}} = [(k _2 -1)/k _2]\sigma _{x, \eta, 0}$ from the
motionless instantaneous orbit $x _{\eta _{3}}$ of ions having
minimum energy and zero amplitudes of betatron oscillations when the
laser beam is stopped at the position $x _{\eta _{3}}$.

If we take into account that the instantaneous orbits of ions having
maximum amplitudes of betatron oscillations and minimum energy are
below the instantaneous orbits of ions having zero amplitudes of
betatron oscillations and minimum energy, by the value $0.28 \sigma
_{x,b,0}$, at the moment of the laser beam stopping then the total
radial dispersion of the instantaneous orbits of the beam can be
presented in the form

         $$\sigma _{x,\varepsilon,f} \leq
         {k_2 - 1\over k _2} \sigma _{x,0} +
         [\sqrt {k _2\over k_2 - 1} - \pi (k _{2} -1)
         \psi (k _2, \xi _{2,f}) $$
         \begin{equation}               
         + 0.28] \sigma _{x, b,0},
         \hskip 6mm A _{T _2} > l _f, \sigma _{x,0}.
         \end{equation}

The damping time of the ion beam in the longitudinal plane $ \tau
_{\varepsilon} = 2 \sigma _{\varepsilon,0}(1 + \sigma _{x,\,b,0}
/\sigma _{x,\,\varepsilon,0})/\overline P$.

According to (20) the efficiency of the enhanced longitudinal laser
cooling is the higher the less the ratio of the spread of the initial
amplitudes of betatron oscillations to the spread of the instantaneous
orbits of the being cooled ion beam.

According to (16) and (20) the enhanced longitudinal laser cooling can
lead to a high degree of cooling of ion beams in the longitudinal
plane and a much lesser degree of heating in the transverse one.

                     \section {Discussion}

In the nonresonance method of the enhanced transverse laser cooling of
ion beams, according to (13) and (14), the degree of decrease of
betatron oscillations $C _{1} = \sigma _{x,b, 0}/ \sigma _{x, b, f} = A
_0 /A = \sqrt{(1 + |k _1|)/ |k _1|}$ and the degree of increase of the
spread of the instantaneous orbits of the beam is much greater: $D _{1}
\simeq C _{1} ^{2}$. In the nonresonance method of the enhanced
longitudinal laser cooling, according to (16) and (20), there is a
significant decrease in the spread of instantaneous orbits of ions $C
_{2} = \sigma _{x, \varepsilon, 0}/ \sigma _{x, \varepsilon, f}$, and a
much lesser value of increase in the amplitudes of betatron
oscillations: $D _{2} \simeq \sqrt{C _{2}}$. From this it follows that
cooling of ion beams both in the transverse and longitudinal planes, in
turn, does not lead to their total cooling in these planes. We can cool
ion beams either in the transverse or longitudinal planes. To cool an
ion beam both in the transverse and longitudinal planes we must look
for combinations of enhanced nonresonance methods of cooling with other
methods.

In the nonresonance method of enhanced longitudinal laser cooling,
contrary to the transverse one, the degree of longitudinal cooling is
much greater than the degree of heating in the transverse plane. That
is why we can use the emittance exchange between longitudinal and
transverse planes when the RF system is switched on and a
synchro-betatron resonance \cite {robinson} - \cite{kihara} or
dispersion coupling by additional motionless wedge-shaped targets \cite
{oneil, shoch, neufer} are used together with the moving there and back
target $T _2$. In such a way the enhanced two-dimensional cooling of
the ion beam based on the longitudinal laser cooling can be
realized\footnote{For the wedge-shaped target the fast cooling effect
in one direction is equal to near the same heating effect in the other
one. That is why the combination of the fast emittance exchange with
the enhanced longitudinal cooling can lead to the two-dimensional
enhanced cooling effect. Cooling of muon beams by material targets can
be done similar way.}. In this case the energy losses of ions in the
laser beam per turn must be higher than the maximum energy gain in the
RF system and the motion of the target $T _2$ must be limited by the
region $x _{T_2} > 0$. Cooling of particle beams in the RF buckets is
another problem to be considered elsewhere.

The enhanced nonresonance transverse method of laser cooling together
with the enhanced resonance longitudinal one can be used for cooling of
ion beams when the RF system is switched off. A heating of the ion beam
in the longitudinal plane in the process of the enhanced transverse
cooling will be compensated completely by its following cooling in the
longitudinal plane.

                    \section{Appendix}

We start by reviewing very briefly the derivation of Robinson's damping
criterion \cite {robinson}.

The general method of describing the motion of a particle in a circular
accelerator is to determine an equilibrium orbit, and then analyze
small deviations from this orbit as a linear combinations of normal
modes of oscillation. The characteristics of the modes are determined
by solving for the principal values of the matrix. If the particle
motion is stable such that the particle oscillate about the equilibrium
position, and since the transfer matrix is real, the principal values
will be three pairs  of complex conjugate numbers, which determine the
frequencies and damping rates of the oscillation modes.

To find damping rates K.W.Robinson considered an element of the
accelerator of infinitesimal length and calculated the six order
transfer matrix for this element. This matrix has infinitesimal
nondiagonal terms which are first order in the length of the element,
and the diagonal terms differ from unity by a quantity which is
proportional to the infinitesimal length of the element. In order to
determine damping, the determinant of the transfer matrix of the
infinitesimal element was evaluated. The only terms in the determinant
which are first order in the length of the element are due to the
diagonal terms of the matrix.  The determinant of the transfer matrix
is given by $1 + \Sigma \delta _{mn}$. The diagonal terms for $x,y,z$
are zero as changes in $x,y$ are only related to $x ^{'},y ^{'}$ and
changes in $z$ related to $x$. In this case $x,x ^{'},y,y ^{'}$
represent the variation of displacement and angular deviation in the
radial and vertical planes, $\Delta \varepsilon$ and $z$ represent the
variation in energy and azimuthal position from the values of an
equilibrium particle, as measured at the time the particle transverses
the infinitesimal element.

The diagonal term for $\Delta \varepsilon$ was determined from the
characteristics of the radiation loss $P _{\gamma} \sim E ^2 B ^2$,

         \begin{equation}               
         P _{\gamma} = P _{\gamma, \,s} (1 + 2\Delta B/B + 2\Delta
         \varepsilon/\varepsilon), \end{equation}
where $B$ is only function of position.  The diagonal term for $\Delta
\varepsilon$ due to radiation loss is $1 - 2\delta \varepsilon _{loss}/
\varepsilon _s$ with $\delta \varepsilon _{loss}$ the radiation loss
for an ideal particle in the infinitesimal element.  The energy gain
from the RF system is not dependent on $\Delta \varepsilon$ and
contributes no change in the $\Delta \varepsilon$ diagonal term.

The difference from unity of $ x ^{'}$ and $ y ^{'}$ diagonal terms is
determined from the energy gain from the RF system and is unaffected by
radiation loss. The energy increase due to the RF system add a momentum
change parallel to the equilibrium orbit and will reduce the angular
variation for the value $\delta x ^{'} = - (\delta \varepsilon _{RF}/
\varepsilon _s) x ^{'}$ and the diagonal term for  $x ^{'}$ is $1 -
\delta \varepsilon _{RF}/ \varepsilon _s$, with $\delta \varepsilon
_{RF}$ the energy gain from the RF system for an equilibrium particle.
Similarly the diagonal element for  $y ^{'}$ is $1 - \delta \varepsilon
_{RF}/ \varepsilon _s$. Then the determinant for the infinitesimal
element is

         \begin{equation}               
         D ^{inf} = 1 + \Sigma \delta _{nn} = 1 - 2 \delta \varepsilon
         _{loss}/ \varepsilon _s -  2 \delta \varepsilon _{RF}/
         \varepsilon _s.
         \end{equation}

The determinant of the transfer matrix for one complete period is the
product of the transfer matrices of the infinitesimal elements of that
period

         \begin{equation}               
         D  = 1 + \Sigma \delta _{nn} = 1 - 2 \varepsilon
         _{loss}/ \varepsilon _s -  2 \varepsilon _{RF}/
         \varepsilon _s,
         \end{equation}
where $\varepsilon _{loss}$ and $\varepsilon _{RF}$ are the radiation
loss and energy gain from the RF system in one period.

The characteristics of the principal modes of oscillation are
determined by solving for the principal values of the transfer matrix
for one complete period. If all modes are oscillatory the principal
values will be of the form $\exp \gamma _i$. Then $D = 1 - 2
\varepsilon _{loss}/ \varepsilon _s -  2 \varepsilon _{RF}/ \varepsilon
_s = \Pi \exp \gamma _i$ or $\exp \Sigma 2 \alpha _i = 1 - 2
\varepsilon _{loss}/ \varepsilon _s - 2 \varepsilon _{RF}/ \varepsilon
_s$, where $\gamma _i = \alpha _i \pm i \nu _i$, $\alpha _i$ is the
fractional damping of a mode in one period of the accelerator.

For equilibrium conditions the radiation loss is equal to the energy
gain from the RF system for one complete period, $\varepsilon _{loss} =
\varepsilon _{RF}$. Then for $\Sigma \alpha _i | _{|\alpha _i| \ll 1} =
- 2 \varepsilon _{loss}/ \varepsilon _s$.

From this it follows the Robinson's damping criterion for the sum of
the damping rates of the three modes of oscillation

         \begin{equation}               
         \Sigma \beta _i = - 2 P _{\gamma \,s}/ \varepsilon _s,
         \end{equation}
where $P _{\gamma \,s}$ is the average rate of radiation loss, and the
amplitude of an oscillation varies as $\exp (\beta _i t)$.

When the vertical and radial oscillations are not coupled then damping
decrements for vertical betatron and synchrotron oscillations can be
derived lightly and the decrement for the radial betatron oscillations
cam be derived from the Robinson's damping criterion. The decrement for
the radial betatron oscillations cam be derived directly as well.

Below we would like to pay attention on the limits of applicability of
this criterion.

As a starting point for his proof K.W.Robinson took some assumptions.
He wrote:

1. "For small deviations from the principal orbit, a transfer matrix
for a complete period may be written relating initial to final
deviations. This is usually done for radial and vertical displacements
and velocities, and may be extended to sixth order transfer matrix
relating initial and final vertical displacements and velocities, and
also energy variation, and longitudinal displacement, from the values
of a particle on the principal orbit. For this general transfer matrix,
the complete periods of the accelerator are defined so as to be
identical in both magnet structure and RF accelerating
system."

2. "The diagonal term for $\Delta \varepsilon$ may be determined from the
characteristics of the radiation loss for small
variations in $E$ and $B$ from the values for an ideal particle. $B$ is
only a function of position, ... ." (see eq. (21).

3. "If all modes are oscillatory the principal values will be of the
form $\exp \gamma _i$, with six values of $\gamma _i$ being three pairs
of complex conjugates."

4. "... This is a general result for any type of electron accelerator
if the average electron energy is constant." (He has in view (24) for
the result).

We can see that the Robinson's damping criterion was received for the
stationary conditions when the guiding magnetic field and the amplitude
of the RF voltage does not depend on time, particle energy loss is a
linear function of both the energy deviation of the particle $\Delta
\varepsilon$ and the magnetic field deviation $\Delta B$ in the region
occupied by the beam, all modes are oscillatory.

Linear dependence on $\Delta \varepsilon$, $\Delta B$ and stationary
conditions are important for the exponential decay and the concept of
decrement. The equations of motion in our case are the linear
homogeneous Hill's equations for the transverse degrees of
freedom and the pendulum equation for the longitudinal one when we
neglect coupling and losses. The losses lead to additional terms at the
first order derivatives in the corresponding equations.

The violation of the Robinson's assumptions can lead to the violation
of his damping criterion. Simple examples can justify this statement.

{\bf Example 1.} Laser cooling of ion beams in the longitudinal plane.
The RF system of the storage ring is switched off. The homogeneous
laser beam overlaps the ion beam. The chirp of the laser frequency is
used.

In this example there is no oscillatory motion of ions in the
longitudinal plane. The instantaneous orbits of ions which are at
resonance with the laser beam are moving with a constant velocity to
the motionless non-resonance orbit of ions having minimal energy. When
the orbit of resonance ions reach the orbit of ions having minimal
energy the laser beam is switched off.  We have non-exponential low of
bringing closer of resonance and non-resonance ions. The damping time
is determined by a value (2), which is out of the Robinson's damping
criterion. The degree of cooling for this time can be very high and is
determined by the natural line width of the laser beam and the maximal
energy of scattered photons.  Heating of the ion beam in the transverse
plane will not appear when the homogeneous laser beam is used.

This is the simplest example of the violation of the Robinson's damping
criteria.  Experiments confirm this conclusion \cite {channel} - \cite
{hangst1}.

We can use the broadband laser beam without chirp and with sharp
frequency edges. Ions in this case will be gathered at the orbit
corresponding to the greatest frequency of the laser beam.

The violation of the Robinson's damping criterion in this example is
the consequence of the non-linear dependence of the power of the
emitted radiation on the deviation of the energy of interacting
and non-interacting ions $\Delta \varepsilon$: ions of the minimal
energy and less does not interact with the laser beam and keep their
position. At the same time ions interacting with the laser beam have
equal velocities. The rate of bringing closer of the energy of
interacting and non-interacting ions is maximum. At the same time the
heating in the transverce direction is absent.

{\bf Example 2.} Laser cooling of ion beams in the longitudinal plane.
The RF system of the storage ring is switched on. The homogeneous laser
beam overlap the ion beam. The broadband laser beam without chirp and
with sharp frequency edges is used. The lowest frequency of the laser
beam corresponds to the equilibrium energy of ions. The power of
radiation scattered by ions of the energy greater then equilibrium one
$\overline P|_{\varepsilon > \varepsilon _s} > 0$ and in the opposite
case $\overline P|_{\varepsilon \leq \varepsilon _s} = 0$.

In this example the power $\overline P(\varepsilon)$ is not linear
function of the ion energy. If the power of scattered radiation is much
higher than the maximum power which the ion can extract from the RF
system then all ions of the energy higher then equilibrium one will be
gathered at the equilibrium energy for a short time (much less then the
period of phase oscillation). Then we can switch off the laser beam,
wait a quarter of the period of phase oscillation, switch on laser beam
again and wait for damping of the next part of the beam which appeared
in the region of the energy $\varepsilon > \varepsilon _s$.  Such beam
manipulations we can repeat  three times. The beam will be cooled in
the bucket in the longitudinal plane. Heating of the ion beam in the
transverse plane will not appear in this case.

In this paper we have considered another, more complicated schemes of
cooling of particle beams based on nonlinear interaction of moving
target with the being cooled beam.

\addcontentsline {toc} {section} {\protect\numberline {6 \hskip 2mm
References}}


\begin{thebibliography}{9}

\bibitem{idea}        
E.G.Bessonov, Proc. Internat. Linear Accel. Conf. LINAC94,
Tsukuba, KEK, 1994, Vol.2, pp.786; Journal of Russian
Laser Research, 15, No 5, (1994), p.403.

\bibitem{prl}                              
E.G.Bessonov and Kwang-Je Kim, Preprint LBL-37458 UC-414, 1995;
Phys. Rev. Lett., 1996, vol.76, p.431.

\bibitem{pac}                              
E.G.Bessonov,  K.-J.Kim, Proc. 5th European Particle Accelerator
Conference, Sitges, Barcelona, 10-14 June 1996, v.2, p. 1196.

\bibitem {wiedemann} 
H.Wiedemann, Particle Accelerator Physics I \ and II (Springer-Verlag,
New York, 1993).

\bibitem {ICFA01} 
E.G.Bessonov, physics/0202040.

\bibitem{channel} P.J.Channel,  
J. Apply Physics, v. 52(6), p.3791 (1981).

\bibitem{selvo} P.J.Channel,  
L.D.Selvo, R.Bonifacio, W.Barletta, Optics Communications, v.116,
(1995), p.374.

\bibitem{shroder}                 
S.Shr{$\ddot o$}der, R.Clein, N.Boos, et al., Phys. Rev.
Lett., v. 64, No 24, p.2901 (1990).

\bibitem{hangst}     
J.S.Hangst, M.Kristensen, J.S.Nielsen O.Poulsen, J.P. Schifter, P.Shi,
Phys. Rev. Lett., v. 67. 1238 (1991).

\bibitem{hangst1}     
J.S.Hangst, K.Berg-Sorensen, P.S.Jessen, et al., Proc. IEEE Part. Accel.
Conf., San Francisco, May 6-9, NY, 1991, v.3, p.1764.

\bibitem{oneil}                     
O'Neil G., Phys. Rev., 102, 1418 (1956);

\bibitem{shoch}                     
A.Shoch, Nucl. Instr. Meth, v.11, p.40 (1961).


\bibitem{robinson}                  
K.W.Robinson, Phys. Rev., 1958, v.111, No 2, p.373.

\bibitem{hoffman}                  
A.Hoffman, R.Little, J.Peterson, Proc. VI Int. Conf. High
Energy Accel. Cambridge (Mass.), 1967, p.123.

\bibitem{sessler}                   
H.Okamoto, A.M.Sessler, and D.M$\ddot o$hl, Phys. Rev.
Lett. {\bf 72}, 3977 (1994).

\bibitem{kihara}                   
T.Kihara, H.Okamoto, Y.Iwashita, et al., Phys. Rev. E, v.59, No 3, p.
3594, (1999).

\bibitem{neufer}                    
D.V.Neufer, Nucl. Instr. Methods, 1994, v.A350, p.24.

\end{thebibliography}
\end{document}